%% file: smc4bio.tex
\documentclass[11pt]{article}
\usepackage[english]{babel}
\usepackage{amsmath, amsfonts, amssymb, amsthm}
\usepackage{stmaryrd, bbm}
\usepackage[ruled]{algorithm2e}
\usepackage{url}
\usepackage{graphicx}
\usepackage{multirow}

\newcommand{\hide}[1]{}
\newcommand{\eg}{{\em e.g.}}
\newcommand{\ie}{{\em i.e.}}
\newcommand{\etc}{{\em etc.}}

\theoremstyle{plain}

\theoremstyle{definition}

\theoremstyle{definition}

\newcommand{\reals}{\mathbb{R}}
\newcommand{\nnreals}{\reals^{\geqslant 0}}

\newcommand{\nats}{\mathbb{N}}
\newcommand{\ints}{\mathbb{Z}}

\newcommand{\ste}{{\bf x}}
\newcommand{\rvec}{{\bf v}}

\title{Statistical Model Checking\\for Biological Applications}
\author{Paolo Zuliani\\School of Computing Science\\Newcastle University\\\url{paolo.zuliani@ncl.ac.uk}}
	
\date{}

\begin{document}
\maketitle

\begin{abstract}
In this paper we survey recent work on the use of statistical model checking techniques
for biological applications. We begin with an overview of the basic modelling 
techniques for biochemical reactions and their corresponding stochastic simulation algorithm - 
the Gillespie algorithm. We continue by giving a brief description of the relation between 
stochastic models and continuous (ordinary differential equation) models. Next we present a
literature survey, divided in two general areas.
In the first area we focus on works addressing verification of biological models, while in the 
second area we focus on papers tackling the parameter synthesis problem.
We conclude with some open problems and directions for further research.
\end{abstract}

\section{Introduction}
\input{intro}

\section{Biological Modelling}
\input{biomodel}

\section{Verification of Biological Models}
\input{verif}

\section{Parameter Estimation and Synthesis}
\input{synthesis}

\section{Tool support}
\input{tools}

\section{Conclusions}
\input{conc}

\bibliographystyle{abbrv}
\bibliography{smc4bio}

\end{document}

%% file: intro.tex
In this paper we survey recent works on the application of statistical model checking
to stochastic biological models. 
Statistical model checking (SMC) is particularly apt for computational analysis of stochastic 
biological models. The reason is that such models are predominantly finite-state (or countably
infinite) continuous-time Markov chains. A
walk, or simulation, through the chain is thus a legitimate trajectory of the {\em actual} stochastic 
process, without any information loss except, of course, for the use of pseudo-random numbers. This
contrasts with continuous models (\eg, ordinary differential equations), where one is forced 
to discretise both time and state space. This can lead to loss of detail in the simulation of 
the process, which in turn can lead to missing certain events in the state space --- this is the so-called 
zero-crossing problem \cite{Zhang08Z}. 

Another reason for the use of SMC in Biology is that most biological models 
are lacking many key parameters, \eg, reaction rate constants. Such parameters must be obtained 
through expensive and time-consuming wet-lab experiments, hence their scarcity. Also, the actual 
value of the parameters in a given
experiment can be affected by imprecisions in the experiment's initial conditions, such as temperature,
reactant concentrations, \etc~Overall, these two hurdles cause most biological models to be
``quantitatively imprecise'' and thus unable to give accurate predictions of the actual behaviour of
the system under study. Hence, it appears to be wasteful to use very precise numeric model checking 
techniques on models that are rough approximation by construction. Also, numeric techniques tend
to scale poorly with the model size and can quickly become unfeasible, thereby leaving simulation 
as the only option. In this work we focus on SMC for stochastic biological models. A general 
introduction to SMC can be found elsewhere in this special issue.
For a broad perspective on general model checking techniques for biological systems, 
we refer the reader to Brim {\em et al.}'s comprehensive recent review \cite{BrimSFM13}. 

The structure of this paper is as follows. First we give an overview of biochemical reaction networks
modelling, focusing on the most used simulation approach - the Gillespie algorithm. In this part we
also briefly elucidate the relation between stochastic models and ordinary differential equation
models. Next we present our literature survey broadly divided into two areas. In the first area
we present papers addressing the verification of biological models. In the second area we focus of papers
tackling the parameter estimation and parameter synthesis problems. Finally, we conclude with
some open problems and possible directions for future research.

%% file: biomodel.tex

A very large portion of the biological modelling literature addresses the problem of
devising efficient computational models of biochemical reactions.
In this Section we give a brief overview of the most used techniques for such a 
problem --- ordinary differential equations and Gillespie's algorithm \cite{SSA}.

\subsection{Biochemical Reactions}

A biochemical reaction specifies a transformation from a bag of molecular types or
 chemical species (the {\em reactants}) to another bag of chemical species 
(the {\em products}).  For example, the reactions for the well-known Michaelis-Menten 
enzyme kinetics are:
\[\begin{array}{rll}
	S + E &\longleftrightarrow & SE \\
	SE &\longrightarrow  &P+E
\end{array}\]
where $E$ is an enzyme that catalyses the production of $P$ from a substrate $S$.
The top line actually defines a {\em reversible} reaction (note the double arrow): 
we have a forward reaction that produces the complex $SE$ from $E$ and $S$, and we have 
a backward reaction that `splits' $SE$ into $S$ and $E$.

In general, a biochemical reaction has the following shape:
\[
	l_1R_1 + \ldots +l_rR_r\quad \longrightarrow\quad m_1P_1 + \ldots +m_pP_p 
\]
where the $R_i$ ($P_i$)'s denotes the reactants (products), $r$ ($p$) is the number of reactant 
(product) chemical species. The naturals $l_i$ and $m_i$ are called {\em stoichiometries},
where $l_i$ specifies the number of molecules of $R_i$ consumed when the reaction `fires', while
$m_i$ denotes the number of $P_i$ molecules produced. It is clear that a reversible reaction
can be written as two separate reactions of the type just described. Also, it is not assumed
that the $R_i$ and $P_i$ are distinct, \ie, it might as well happen that a given chemical
species is both consumed and produced in a reaction.

\subsection{Differential Equation Models}

In ordinary differential equation (ODE) models we are interested in modelling the (continuous) 
time evolution of the {\em concentration} of each molecular species. The concentration $[S]$ of 
chemical species $S$ is calculated dividing the actual number of molecules of $S$ by the volume 
of the container where the reactions take place. (Concentrations are usually measured in moles 
per litre, $M$.)
In many applications, it is correct to assume that the `speed' or rate of a reaction is directly
proportional to the concentration (hence mass) of each reactant involved raised to the power
of its stoichiometry coefficient. This is known as {\em mass-action kinetics}. For example,
the forward reaction of the Michaelis-Menten model is assumed to proceed, \ie, producing the 
complex $SE$, at an instantaneous speed proportional to $[S][E]$ (note that concentrations 
change continuously with time, although the notation usually omits this dependency.) In particular,
this also means that $S$ and $E$ are both consumed at a rate $k_f[S][E]$, where $k_f$ is the
{\em rate constant} for the forward reaction and $k_f[S][E]$ is its {\em rate law}.

\begin{figure}[ht!]
        \centering
        \includegraphics[width=13cm]{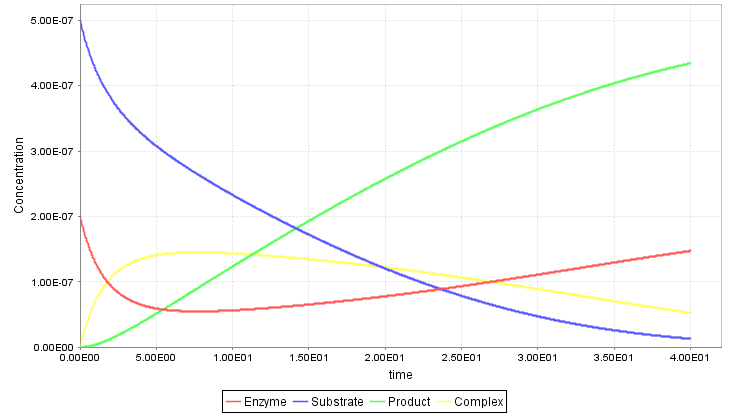}
	\caption{Numerical solution of the Michaelis-Menten ODE system (\ref{def:MMODE}); rate 
	constants: $k_f = 10^6, k_b = 10^{-4}$, and $k_p=0.1$; initial concentrations: 
	$E={\bf 2\times 10^{-7}}, S={\bf 5\times 10^{-7}}, P=C=0$.}
	\label{fig:MMODE}
\end{figure}

To illustrate the use of differential equations, we write the Michaelis-Menten model with the
rate constants:
\begin{equation} \label{def:MM}
\begin{array}{rll}
	S + E &\overset{k_f}{\longrightarrow} & SE \\
	SE &\overset{k_b}{\longrightarrow} & S+E \\
	SE &\overset{k_p}{\longrightarrow}  &P+E
\end{array}
\end{equation}
and assuming mass-action kinetics we can write the following four differential equations, one
for each species:
\begin{equation}\label{def:MMODE}
\begin{array}{rll}
	[S]^\prime &=& k_b[SE] - k_f[S][E] \\[1ex]
	$[$E$]$^\prime &=& (k_b+k_p)[SE] - k_f[S][E] \\[1ex]
	$[$SE$]$^\prime &=& k_f[S][E] - (k_b+k_p)[SE] \\[1ex]
	$[$P$]$^\prime &=& k_p[SE] 
\end{array}
\end{equation}
Again, recall that the concentrations are functions of time. Once the rate constants 
$k_f, k_b$, $k_p$, and the initial concentrations are specified, then the dynamics of the
model is fully specified. The resulting ODEs can be solved analytically only in very few cases,
so the solution is most often obtained via numerical computation.
In Figure \ref{fig:MMODE} we plot a snapshot of the numerical solutions of the ODE 
system (\ref{def:MMODE}) for a particular set of initial concentrations and rate constants.

\subsection{The Stochastic Simulation Algorithm}
Consider a system made of $M$ chemical reactions $R_1,\ldots,R_M$ and $N$ chemical species.
The state of the system at time $t$ is given by the vector 
$\ste(t) \in \nats^N$, {\em i.e.}, the state is just the {\em number of molecules} for each species.
(Again, we drop time dependency and just write $\ste$.)
The state of the system changes only when a reaction fires. Reaction $R_j$ changes the state 
from $\ste$ to $\ste+\rvec_j$, where $\rvec_j \in \ints^N$ denotes the (fixed) state change 
caused by $R_j$. Therefore, we must have $M$ state change vectors $\rvec$'s, one for each reaction.
Note that the elements of a vector $\rvec_j$ are negative integers when molecules of a 
particular species are consumed, while positive integers denote production of species.

The fundamental element in the Stochastic Simulation Algorithm (SSA) \cite{SSA} is the 
{\em propensity} function. To each reaction $R_j$ there corresponds a function 
$a_j:\nats^N\rightarrow \nnreals$, which describes a probability law as follows:
\[
	\text{Prob}(R_j\; \text{fires in interval} \; [t,t+ dt)\, |\, \ste) = a_j(\ste) dt\ .
\]
Note that the propensity functions do not explicitly depend on time (but remember that $\ste$ 
is the state at time $t$).

It can be shown that the (infinitesimal) probability that the system evolves
away from $\ste$ in the time interval $[t+\tau, t+\tau+d\tau)$ is
\begin{align}\label{eq:timedensity}
	\text{Prob}(\text{system evolves}\  \text{in}\; 
	[t+\tau, t+\tau+d\tau)\, |\, \ste) = a_0(\ste) e^{-a_0(\ste)\tau} d\tau
\end{align}
where $a_0(\ste) = \sum_{j=1}^M a_j(\ste)$. Therefore, the time of the next reaction is a random
variable distributed as an exponential with mean $\frac{1}{a_0(\ste)}$.
Now, the probability that system actually chooses reaction $R_j$ is
\begin{align}\label{eq:reactselect}
	\text{Prob}(R_j \ \text{fires}\, |\, \ste) = \frac{a_j(\ste)}{a_0(\ste)}
\end{align}
thus a discrete random variable.
Therefore, by (conditional) event independence we multiply (\ref{eq:timedensity}) and 
(\ref{eq:reactselect}) to get 
\begin{align}\label{eq:prob}
	\text{Prob}(R_j\  \text{fires in}\; [t+\tau, t+\tau+d\tau)\,|\, \ste) 
	 = a_0(\ste) e^{-a_0(\ste)\tau} d\tau \cdot \frac{a_j(\ste)}{a_0(\ste)}\ .
\end{align}
The SSA thus simulates the evolution of the system by sampling two random variables: it samples
the time of the next reaction according to the density (\ref{eq:timedensity}), and it selects which
reaction to fire by sampling from the distribution (\ref{eq:reactselect}). The simulation stops when
a given time bound $T$ is reached (see Algorithm \ref{alg:SSA}). 

\begin{algorithm}[h]
\label{alg:SSA}
\tcp{initialise time and system state}
$\tau$ := 0\;
$\ste := \ste_0$\;
\tcp{simulate up to time $T$}
\While{$\tau \leqslant T$}{
	evaluate $a_j(\ste)$ ($1\leqslant j \leqslant M$) and $a_0(\ste)$\;
	$t := \text{sample time step from density of Eq.}$ (\ref{eq:timedensity})\;{\label{SSAt_i}}
	$j := \text{sample reaction index from distribution of Eq.}$ (\ref{eq:reactselect})\;{\label{SSAr_i}}
	$\tau := \tau + t$\;
	$\ste := \ste + \rvec_j$\;
}

\caption{The Stochastic Simulation Algorithm}
\end{algorithm}

From (\ref{eq:timedensity}) and (\ref{eq:reactselect}) we see that the propensity functions
actually define a time-homogeneous Continuous-Time Markov Chain (CTMC), whose state space can possibly be 
(countably) infinite.  For finite-state CTMCs, it can be shown that during a finite simulation
({\em i.e.}, $T<\infty$) the state space explored is finite with probability one. 
That is, the state transition sequences of finite CTMCs satisfy the non-zenoness property \cite{BHHK03}. 
In our setting, this means that the SSA will 
fire an infinite number of reactions in finite time with probability zero.

To complete the description of the SSA, we only need to define explicitly the propensity functions.
As for ODE modelling, one may generally assume mass-action kinetics. For example, the 
{\em stochastic rate law} (or propensity)
for a unimolecular reaction such as the Michaelis-Menten production reaction is $c_p{\bf [se]}$, where
$c_p$ is the {\em stochastic rate constant} for the reaction and ${\bf [se]}$ is the quantity of species
$SE$ (remember that ${\bf [se]}$ is a natural number.) For bimolecular reactions such as the forward
reaction, the rate law is $c_f{\bf [s][e]}$, where ${\bf [s]}$ and ${\bf [e]}$ are the quantity of
species $S$ and $E$, respectively. For the case of a general reaction,
and to see how the stochastic rate constants can be obtained from the deterministic (ODE) constants,
the interested reader can find more details in \cite{Wilkinson}, for example.

\begin{figure}[ht!]
        \centering
        \includegraphics[width=13cm]{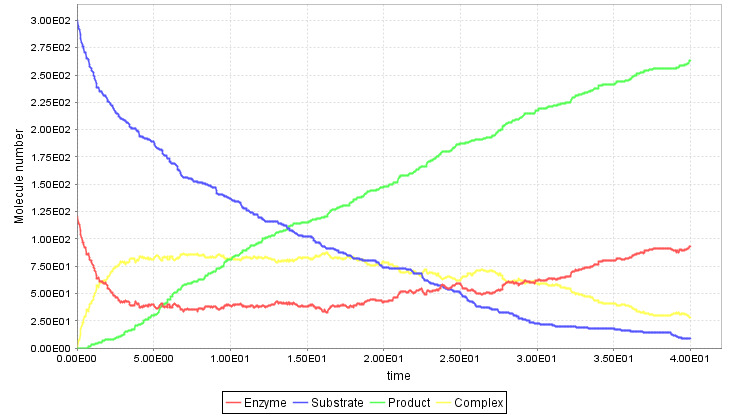}
	\caption{An exemplary run of the SSA on the Michaelis-Menten reaction system (\ref{def:MM}); 
	stochastic rate constants: $c_f = 1.66\times 10^{-3}, c_b = 10^{-4}$, and $c_p=0.1$; initial 
	molecule numbers: $E={\bf 121}, S={\bf 300}, P=C=0$.}
	\label{fig:MMSSA}
\end{figure}

In Figure \ref{fig:MMSSA} we plot an exemplary run of the SSA on the Michaelis-Menten set of
reactions (\ref{def:MM}). The initial molecule numbers and stochastic rate constants were converted
from the respective deterministic (ODE) values used for generating Figure \ref{fig:MMODE}.
As we can see from the plot in Figure \ref{fig:MMSSA}, the temporal evolution of the species' quantity 
fluctuates, and it is not a smooth line as in the ODE solution of Figure \ref{fig:MMODE}.
Typically, no two SSA runs are the same, except in the case of large initial molecule numbers --- see
Section \ref{sec:SSAwODE} below.

\begin{figure}[ht!]
        \centering
        \includegraphics[width=13cm]{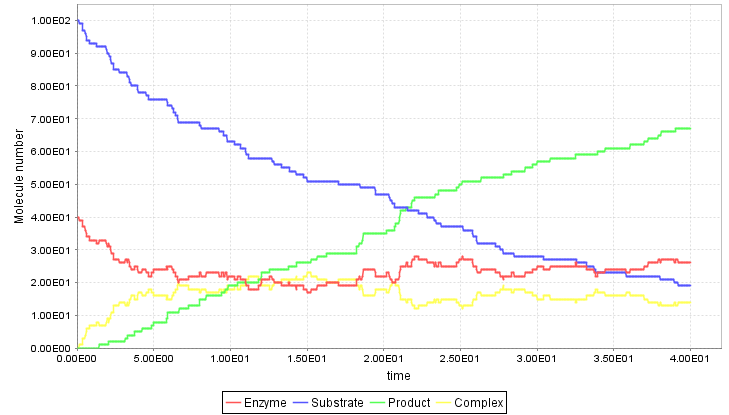}
	\caption{An exemplary run of the SSA on the Michaelis-Menten reaction system (\ref{def:MM}); 
	stochastic rate constants: $c_f = 1.66\times 10^{-3}, c_b = 10^{-4}$, and $c_p=0.1$; initial 
	molecule numbers: $E={\bf 40}, S={\bf 100}, P=C=0$.}
	\label{fig:MMSSAvlown}
\end{figure}

In Figure \ref{fig:MMSSAvlown} we give an example of what happens instead at low molecule numbers:
stochastic effects are even more preponderant than in the setting of Figure \ref{fig:MMSSA}, and 
the temporal evolution of the species seems to be very ``noisy". Also, note that the CTMC nature
of the model is more evident here. In fact, it is easy to see time intervals (\eg, at about 17s) in 
which the species' numbers do not change at all, meaning that no reaction has fired in those intervals.

\subsection{Probability Estimation}\label{SSAestim}

The SSA can be used in a simple Monte Carlo approach to estimate the probability of events 
such as 
\[
	\xi= \text{within $T$ time units the state \ste\  of the system will reach region O}
\]
where $O\subset \nats^N$. As usual, one can define a Bernoulli random variable whose success probability
is Prob($\xi$). Then, the system is repeatedly simulated up to time $T$, while checking at
every step whether $\ste \in O$ (and stopping the simulation if that is true). Finally, to get an 
estimate of Prob$(\xi)$, we divide the number of ``successful'' trajectories - those where $\xi$ occurs -
by the total number of trajectories (see Algorithm \ref{alg:estimSSA}). 
One can use statistical estimation techniques ({\em e.g.}, Chernoff bound) to get confidence intervals for 
the estimate, or use appropriate hypothesis tests.

In the SMC setting, the events $\xi$ are represented by formulae of a suitable temporal logic.
In particular, formulae need to be decidable on a {\em finite} simulation trace, since any
simulation must terminate at some point. An example of such formulae is
\[
	{\bf F}^{25} (\text{\em Product} > 40)
\]
which is true if and only if within the next 25s the amount of {\em Product} will raise above 40. 
Essentially, SMC boils down to either estimating or hypothesis testing the probability of
such properties.

\begin{algorithm}[h]
\label{alg:estimSSA}

$s := 0$\;
\For{$i\leftarrow 1$ \KwTo $K$}{
\tcp{initialise time and system state}
$\tau$ := 0\;
$\ste := \ste_0$\;
\tcp{simulate up to time $T$}
\While{$\tau \leqslant T$}{
	\If{$\ste \in O$}{\tcp{record occurrence of event $\xi$ and stop simulation} $s:=s+1$\; {\bf break}}
	evaluate $a_j(\ste)$ ($1\leqslant j \leqslant M$) and $a_0(\ste)$\;
	$t := \text{sample time step from density of Eq.}$ (\ref{eq:timedensity})\;
	$j := \text{sample reaction index from distribution of Eq.}$ (\ref{eq:reactselect})\;
	$\tau := \tau + t$\;
	$\ste := \ste + \rvec_j$\;
}
}
\Return{$\frac{s}{K}$}

\caption{Probability estimation using the SSA}
\end{algorithm}

\subsection{Relation with Differential Equation Models} \label{sec:SSAwODE}

\begin{figure}[ht!]
        \centering
        \includegraphics[width=13cm]{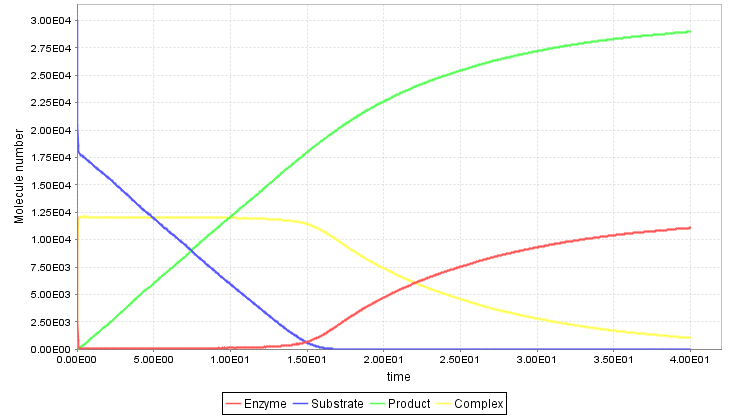}
	\caption{An exemplary run of the SSA on the Michaelis-Menten reaction system (\ref{def:MM}); 
	stochastic rate constants: $c_f = 1.66\times 10^{-3}, c_b = 10^{-4}$, and $c_p=0.1$; initial 
	molecule numbers: $E={\bf 12,100}, S={\bf 30,000}, P=C=0$.}
	\label{fig:MMSSAhn}
\end{figure}

\begin{figure}[ht!]
        \centering
        \includegraphics[width=13cm]{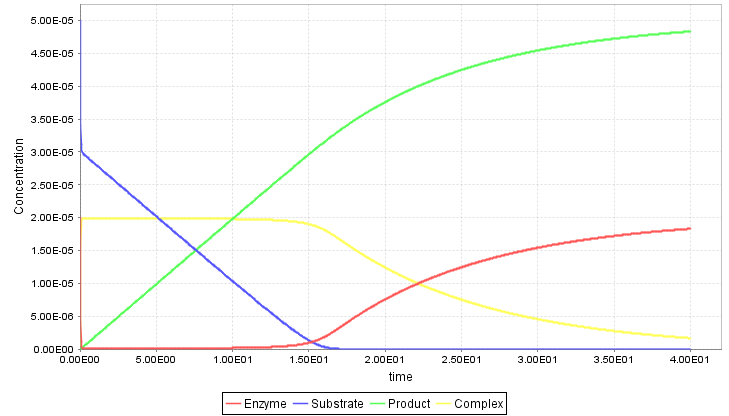}
	\caption{Numerical solution of the Michaelis-Menten ODE system (\ref{def:MMODE}); rate 
	constants: $k_f = 10^6, k_b = 10^{-4}$, and $k_p=0.1$; initial concentrations: 
	$E={\bf 2\times 10^{-5}}, S={\bf 5\times 10^{-5}}, P=C=0$.}
	\label{fig:MMODEhc}
\end{figure}

It can be shown that, for any given finite volume, for high molecule numbers (hence high 
concentrations) the stochastic (CTMC) model can be very well approximated by its corresponding ODE 
model. On the other hand, for low molecule numbers the stochastic and ODE model will in general 
behave very differently. The same effects can be had if we fix the molecule numbers and vary the
volume, of course. In Figure \ref{fig:MMSSAhn} we plot a run of the SSA on the Michaelis-Menten
reactions (\ref{def:MM}) with a 100-fold increase of the initial molecule numbers with respect 
to those used in Figure \ref{fig:MMSSA}.
Notice how the species' numbers evolve almost ``continuously", in a very similar way to the 
solutions of the corresponding ODE system plotted in Figure \ref{fig:MMODEhc}.

From a computational point of view, the cost of simulating a set of ODEs
does not depend much on the initial conditions. However, it is easy to see that for high molecule
numbers, the cost of simulating the CTMC reaction model increases significantly. In particular,
the mean time of the next reaction is inversely proportional to the total number of molecules 
present in the system and the reaction rates --- see Eq.~(\ref{eq:timedensity}). Therefore, 
to simulate a given amount of system time, the SSA will generate a much higher number of reaction 
events. This is the reason why many researchers are still actively investigating ways to speed up 
the SSA while maintaining a reasonable accuracy. The $\tau$-leaping technique is such
an example \cite{Gillespie01,Rathinam03}.

The ODE approximation of CTMCs in the context of biochemical modelling was formally proved in 
a 1992 paper by Gillespie \cite{Gillespie92}, although in fact the general theory had been 
developed in 1970 by Kurtz \cite{Kurtz70}.
The same theory has been recently introduced in performance modelling (but known as fluid flow 
approximation.)

%% file: verif.tex
\subsection{Biochemical Reaction Networks}

The idea of using a statistical (or Monte Carlo) approach for model checking stochastic 
biological models was introduced as early as 2006 by Calzone {\em et al.}~\cite{Calzone06}. 
In that paper, the authors outline a Monte Carlo approach for estimating the probability of 
(a finite variant of) LTL formulae over CTMC models of biochemical reactions, along the lines 
presented in Section \ref{SSAestim}. The authors do not discuss the type of statistical tests 
to be used, although they seem to hint at the Chernoff bound.

In a 2008 work, Clarke {\em et al.}~\cite{Biolab} have applied Wald's Sequential Probability 
Ratio Test (SPRT) \cite{Wald45} for verifying stochastic models of signalling pathways. Properties 
are expressed using a novel formalism, the Bounded Linear Temporal Logic (BLTL). This logic is 
essentially a variant of LTL in which temporal operators are equipped with a time bound. Also, BLTL 
can be seen as a sublogic of Metric Temporal Logic \cite{Koymans90}. Previous approaches used a 
version of LTL defined over finite traces, with the addition of a {\em time} proposition representing 
the `global' time of the simulation. Hence, BLTL offers better flexibility as every temporal operator 
has its own `clock'. We note that the use of the SPRT for statistical model checking was first introduced 
by Younes and Simmons' in their pioneering paper \cite{Younes02}.

A Bayesian approach to SMC was first introduced by Jha {\em et al.}~in a 2009 paper \cite{Jha09}.
In particular, the authors introduce a sequential version of the Bayes factor (hypothesis) test.
It is a test for {\em composite} hypotheses, \ie, $p \leqslant \theta$ {\em vs.} $p > \theta$, where
in our SMC setting $p$ is the unknown probability that a given property holds. Basically, the test
starts with a {\em prior} distribution over $p$, which is then sequentially updated via Bayes' theorem 
to a {\em posterior} distribution computed as a function of the prior and the likelihood of the 
simulated sample. On the other hand, we recall that the SPRT is in fact a test for {\em simple} 
hypotheses, \ie, $p=\theta_1$ {\em vs.} $p=\theta_2$ (where $0\leqslant \theta_1 < \theta_2 \leqslant 1$)
and the so-called {\em indifference region} is the distance $\theta_2-\theta_1$.
It can be shown that for a large class of distributions (including the SMC relevant Binomial) the
SPRT can be used to decide the composite test $p\leqslant \theta_1$ {\em vs.} $p\geqslant\theta_2$, while
respecting the same Type I and II error probabilities of the simple hypothesis test.
The number of samples required by the SPRT is inversely proportional to the square of the size of
the indifference region, while for the Bayes factor test a similar relation holds with respect
to the distance between the threshold $\theta$ and the true probability of the given formula.
Also, it can be shown that the SPRT is in fact the Bayes factor test for simple
hypotheses with equal {\em a priori} probabilities.

In a 2012 paper, Ballarini {\em et al.}~\cite{Ballarini12} have proposed a SMC approach in which 
properties are expressed in Hybrid Automata Stochastic Logic (HASL) instead of one of the usual
variants of temporal logic. Basically, a HASL property is made of a linear hybrid automaton
and an arithmetic expression over the so-called {\em data variables} of the automaton. The
goal of the SMC procedure is to estimate the value of the expression by simulating the
(synchronous) composition of the model and the HASL automaton.  The whole approach has been 
implemented in the COSMOS tool \cite{COSMOS}. The authors demonstrate their
technique by analysing stochastic models of gene networks with delayed dynamics. The introduction
of a time delay is meant to capture more faithfully the actual timings involved in fundamental
gene expression processes, \ie, transcription and translation. A drawback of this approach
is the difficulty of writing HASL specifications that capture a given behaviour,\eg, oscillations.
That essentially amounts to specify an automaton, and hence it might constitute a roadblock
for non-expert users.

Koh {\em et al.}\cite{Koh12} have presented OSM, a new hypothesis test algorithm based on the SPRT,
in which one does not have to specify the indifference region. Also, if this new algorithm
cannot reach a conclusion within an allotted amount of time, the p-value of the two hypotheses
(with respect to the samples seen so far) will be computed and returned to the user, who is then
left with the final choice. The approach is a simple modification of Younes' SPRT version with 
error control in the indifference region \cite{Younes06}. The key modification is a dynamic 
adjustment of the indifference region, from the largest possible value (\ie, 1) to the largest 
value that enable a SPRT decision between the null and the alternative hypothesis. The approach is 
applied for verifying temporal logic properties of a neuron subnet model of the worm {\em C.~elegans}. 
The authors have empirically shown that their algorithm is more efficient than Younes' original algorithm.

In Table \ref{tbl:tests} we summarise the various hypothesis tests mentioned in this Section.

\begin{table}[ht]
\centering

\begin{tabular}{l|c|c}
\hline
\hline
{\bf Test} & {\bf Problem addressed} & {\bf Notes} \\
\hline
\multirow{2}{*}{SPRT \cite{Wald45}} & $H_0:p=\theta_1$ {\em vs.} $H_1:p=\theta_2$ &  Also usable for \\
& where $0 \leqslant \theta_1 < \theta_2 \leqslant 1$ & $H_0:p\leqslant\theta_1$ {\em vs.} $H_1:p\geqslant\theta_2$ \\
\hline
\multirow{2}{*}{Bayes \cite{Jha09}} & $H_0: p \leqslant \theta$ {\em vs.} $H_1:p>\theta$ & Easy to modify for\\
	& where $0<\theta<1$ & $H_0:p\leqslant\theta_1$ {\em vs.} $H_1:p\geqslant\theta_2$ \\
\hline
\multirow{2}{*}{OSM \cite{Koh12}} & $H_0: p \leqslant \theta$ {\em vs.} $H_1:p>\theta$ & Based on the SPRT;\\
	& where $0<\theta<1$ & may return `undecided' \\
\hline
\end{tabular}
\caption{Hypothesis testing techniques used in SMC; $p$ is the (unknown) mean of a 
Bernoulli distribution.}

\label{tbl:tests}
\end{table}

\subsection{Other Biological Models}

In a recent paper, Sankaranarayanan and Fainekos \cite{SankaranarayananF12} have analysed the risks
resulting from malfunctioning of insulin infusion pumps by modelling the insuline-glucose regulatory 
system. The model is essentially a Stateflow/Simulink hybrid system with nonlinear 
differential equations. Randomness is introduced by sampling several initial model parameters, 
such as meal time, pump calibration error, \etc~The author have used Bayesian SMC \cite{ZHSCC10} 
for estimating the probabilities of hyper- and hypoglycemia conditions, which are expressed as 
Metric Temporal Logic formulae. (In Section \ref{sec:paramest} we survey another work on 
the synthesis of safe controllers for insulin infusion pumps.)

Jha and Langmead \cite{JhaL12} have investigated SMC in the context of rare-event behaviour in
biological models expressed as stochastic differential equations (SDEs). The behaviours under study are
expressed as BLTL formulae, while the systems studied are models of tumour progression.
The verification problem is then to decide whether a given property is satisfied
with a probability larger than a threshold --- the problem thus reduces to hypothesis testing.
The technique presented basically uses importance sampling to simulate a higher proportion of the 
`important', \ie, rare, event from the SDE model. Then, a modified version of the Bayes factor test is used 
to decide the verification problem. The authors have also derived bounds on the probability of returning the
wrong answer.

%% file: synthesis.tex
The parameter estimation problem is to determine a combination of model parameters
that fit experimental results (\eg, time-series data) or some well-specified
behaviour (\eg, temporal logic formulae.) Parameter synthesis is similarly defined,
except that one is instead interested in determining {\em sets} of parameters that
satisfy the given constraints. Obviously, parameter synthesis is a much harder
problem than parameter estimation. In the setting involving biochemical reactions, the
parameters in question are the reaction rate constants of the model.

\subsection{Parameter Estimation}\label{sec:paramest}
In a 2006 paper, Calzone {\em et al.}~\cite{Calzone06} suggested that a SMC
approach could be used to learn model parameters from LTL specifications. However, the authors also
argued that such an approach would not be practical because of the complexity of the (stochastic) 
simulation and model checking. Hence, no result for stochastic models is actually reported in the 
paper, although the authors do apply their method to continuous deterministic models \cite{Calzone06}.
Also, note that the procedure could be used for parameter synthesis, since it is essentially a 
simple sweep of the whole (compact) parameter space, up to a certain numerical precision.

Donaldson and Gilbert \cite{Donaldson08} presented in 2008 a SMC approach for estimating reaction
rate constants from temporal logic specifications and numerical constraints. In particular, the
temporal logic used is a variant of LTL that allows for arithmetic expressions and general
functions of the state (\eg, derivative) as atomic properties. The search in the parameter space
is performed by a genetic algorithm whose objective is minimising the distance between the 
behaviour of the current model and the target model. The authors motivate the use of genetic
algorithms because of their ability to avoid getting trapped in local optima. The authors
apply the technique to a large model, and estimate the values of all its 65 parameters.
Even though the model analysed is continuous, we have reported here this work because the authors 
defined their framework to be usable on stochastic models, too.

In a 2012 paper, Jha {\em et al.}~\cite{Jha12} have studied the problem of synthesising control laws
for insulin infusion pumps from safety specifications expressed in BLTL. The glucose-insulin model 
is a relatively simple ODE model, while the controller implements a Proportional Integral Derivative 
control law, described by an equation containing both integration and derivation (of course, the entire 
model can be converted to an equivalent ODE-only model.) The task is to find values for the three parts 
of the control law in such a way that the controller meets its safety specifications. This is 
achieved by an exhaustive sweep of the (bounded) parameter space. Also, the authors assume that some 
of the model's parameters are random, so that the ODE model becomes suitable for SMC.

Recently, Palaniappan {\em et al.}~\cite{Palaniappan13} have introduced a SMC-based approach for
calibrating ODE models of signalling pathways. In particular, the authors present a technique for
parameter estimation with respect to behavioural specification and experimental data that are
described as BLTL formulae. The idea is quite straightforward: the unknown parameters are sampled 
from (usually uniform) distributions, the resulting ODE model is then simulated, and finally the SPRT 
is used to decide whether the given probabilistic BLTL formula is satisfied or not. An evolutionary
algorithm is used to guide the search in the parameter space, through a goodness-of-fit function that
measures how well the current model agrees with the behavioural specification and the experimental data.
The same function is also used to perform sensitivity analysis. The authors have successfully applied 
their approach to a reasonably large (105 ODEs, 39 unknown parameters) pathway model.

In the past year, verification and statistical model checking have started embracing the notion of
{\em robustness} for temporal logic. The idea is to endow temporal logic with a suitable 
{\em quantitative} semantics that: 1) generalises the standard (Boolean) semantics; and 2) gives
a measure of ``how much'' a temporal formula is true. For example, consider the formula 
${\bf F} (x > 0)$: intuitively, a trace would satisfy the formula more robustly if the {\em maximum} value
of $x$ in such a trace is higher (conversely, if $x$ is always negative then the formula robustness 
will be negative, and the Boolean semantics is false, of course). Donz\'{e} {\em et al.}~\cite{DonzeFM13}
defined a quantitative semantics for STL (a generalisation of LTL in which the Until operator may 
be equipped with an interval) and gave algorithms for computing the robustness of a STL formula 
with respect to time-series data (or traces).

Bartocci {\em et al.}~\cite{BartocciBNS13} have extended the notion of STL robustness to stochastic 
(CTMC) models. In this setting, robustness becomes a random variable defined on the space of the
stochastic model's traces. Also, one obtains the robustness {\em distribution}, \ie, the probability
distribution associated to the robustness random variable. For example, the mean of the robustness
distribution gives the `average' robustness of the model traces. The authors use the
robustness distribution to guide parameter estimation for CTMC models, \ie, find a parameter
combination such that the model satisfies a given STL formula with probability greater than a given
threshold, while {\em maximising} its robustness. In particular, statistical model checking 
(on STL properties) is used for building a statistical emulator of the robustness distribution, upon
which an optimisation algorithm is then used to identify the parameter combination. The use of a 
statistical emulator is motivated by the fact that evaluating the true robustness distribution is
computationally onerous.

\v{C}eska {\em et al.}~\cite{Ceska14} explore another definition of robustness, based on an earlier
notion by Kitano \cite{KitanoRobust}. Basically, they define robustness as the expected value of an
{\em evaluation function} with respect to a probability measure over a space of system perturbations. 
The evaluation function is defined over the perturbation space, and it determines to what degree 
the ``wild-type'' behaviour of the system is preserved by a given perturbation. The evaluation
function is based on the probability that the model satisfies a given CSL formula. To compute the 
expectation, the authors employ a novel probabilistic model checking technique that can provide 
bounds on the evaluation function, since computing the function for every perturbation point is 
unfeasible. Even though the current approach is based on probabilistic model checking, the authors 
report that they are looking into extending it towards statistical model checking in order to cope 
with dimensionally large parameter spaces.

\subsection{Parameter Synthesis}\label{sec:paramsyn}
In \cite{JhaTCS11}, Jha and Langmead present three algorithms for synthesising kinetic constants
of stochastic (CTMC) reaction models given a probabilistic temporal logic specification, a parameter
search region (a multi-dimensional real compact set), and an error tolerance. The algorithms decide
whether there are parameters in the region that satisfy the specification (up to the given error) 
or whether the model is {\em unfeasible} within the given parameter region. The difference between
the three algorithms lies in the output: two algorithms return, if it exists, a subregion of the 
parameter search space for which every point is guaranteed to satisfy the probabilistic temporal 
formula --- so called {\em feasible} region. (The second algorithm uses an abstraction-refinement 
technique to speed up convergence.) The third algorithm returns instead the point in the feasible 
region that maximises the probability that the temporal logic formula is true, thereby solving
parameter estimation.

The algorithms rely on the fact that the probability density in the SSA is a continuous function 
of the kinetic parameters, and thus uniformly continuous as the parameter region is a compact set.
Therefore, the probability that the formula is true can be bounded accordingly on any subset
of the parameter region. This is the key result that enable the parameter synthesis algorithms:
basically, one first partitions the parameter region according to the error tolerance selected
by the user, and then one point is randomly sampled in each partition sets. If the point satisfies 
the formula then the entire partition set is added to the feasible region, otherwise that partition
set is rejected. Note that the size of each partition set is computed in such a way that for any point
in a given partition set the probability of the formula being true does not change more than 
the error tolerance. The third (parameter estimation) algorithm works in a similar way except 
that it uses the gradient of the satisfaction probability of the formula to guide the search 
in the parameter region. The authors successfully apply their parameter synthesis techniques 
to a six-dimensional biological model, while their parameter estimation algorithm was able 
to scale up to eleven parameters. Finally, we remark that all the results given by the algorithms 
are necessarily approximate, because of general undecidability results for complex first order
formulae over the reals \cite{Richardson68,GaoLICS12}.

%% file: tools.tex
In this Section we give a brief overview of software tools that support SMC, which are
now in growing numbers. We have the aforementioned COSMOS tool \cite{COSMOS} for SMC 
of stochastic hybrid automata, which has been applied for studying gene networks.
The ANIMO tool \cite{ANIMO} allows the user to build signalling pathway models in a 
graphical way. The models can be simulated and also verified using the UPPAAL
model checker \cite{UPPAAL}, since the underlying semantics is given as timed automata.

PLASMA-Lab \cite{PlasmaLab} is a very flexible library for embedding statistical 
model checking capabilities in general simulation tools. In particular, PLASMA-Lab can 
verify biological models expressed as CTMCs via its own modelling language and simulator; 
the property language is BLTL extended with several temporal operators for minimum, 
maximum and mean of a variable. PLASMA-Lab can also efficiently handle the verification 
of rare events, \ie, properties that are satisfied with extremely low probability.

PRISM \cite{PRISM} is a very powerful model checker that can handle a large variety
of stochastic models (discrete-time MC, CTMC, Markov Decision Processes, Probabilistic 
Automata, {\em etc}.) and temporal logics (CSL, LTL, {\em etc.}). For our application area, 
PRISM's modelling language can directly support CTMCs, and temporal logic verification 
can be performed either numerically or via SMC. The former is very precise, but it
can become quickly infeasible for large models. PRISM has been used for many case
studies in systems 
biology\footnote{\url{http://www.prismmodelchecker.org/casestudies/index.php#biology}}. 

UPPAAL-SMC \cite{UPPAALSMC} extends UPPAAL by adding SMC capabilities for networks of 
stochastic hybrid automata. The modelling language is quite rich, allowing the user 
to define hybrid models containing both CTMCs and ODEs. UPPAAL-SMC has been used for 
studying a number of biological case studies, \eg, a genetic oscillator \cite{DavidLLMPS12}.
Also, UPPAAL-SMC supports SBML models (see below) through an internal conversion utility.

Finally, the Systems Biology Markup Language (SBML) \cite{SBML2003} is a very much used 
language for interchanging computational biology models. It is supported by most tools
for scientific computation such as, \eg, \hyphenation{MATLAB} MATLAB (via the SimBiology 
package or SBMLToolbox \cite{SBMLToolbox}). A list of SBML compliant tools is available on
\url{http://sbml.org}.

%% file: conc.tex
Statistical model checking (SMC) is becoming increasingly useful for analysing stochastic biological 
models, as witnessed by a number of recent works surveyed in this paper. The use of SMC for `traditional'
verification of biological models appeared first, and it is perhaps now the only way to tackle the
usually large and complex models arising from biological applications. Now, SMC is being investigated 
for other difficult problems, such as parameter estimation and parameter synthesis, which are even 
more central to systems biology. It is expected that computer science researchers will make use of the
rich literature on parameter estimation and synthesis from other disciplines, \eg, control theory.
The estimation of rare-event probabilities for biological models might represent another area for
further work. It is well known that standard SMC would require extremely large sample sizes to estimate 
accurately very low probabilities. Finally, a whole new area of possible application for SMC techniques
is represented by sequence analysis. In that field, hidden Markov models are routinely used to analyse
DNA sequences, and one could conceivably use some variant of temporal logic to express sequence properties
and then apply SMC for verification.